\documentclass{article}

\usepackage{graphicx}

\usepackage{amssymb}

\begin{document}

%\begin{frontmatter}



\title{A Study of the Cyclotron Gas-Stopping Concept for the Production of Rare Isotope Beams}

\author{M. Sternberg, G. Savard \\
	Argonne National Laboratory and The University of Chicago}

\maketitle

%\address[argonne]{Physics Division, Argonne National Laboratory, Argonne, IL 60439, USA}
%\address[UofC]{Department of Physics, University of Chicago,
%Chicago, Illinois 60637, USA}

\begin{abstract}
The proposed cyclotron gas-stopping scheme for the efficient thermalization of intense rare
isotope beams is examined. Simulations expand on previous studies
and expose many complications of such an apparatus arising from
physical effects not accounted for properly in previous work. The
previously proposed cyclotron gas-stopper geometry is found to
have a near null efficiency, but extended simulations suggest that
a device with a much larger pole gap could achieve a stopping
efficiency approaching roughly $90\%$ and at least a 10 times larger
acceptance. However, some of the advantages that were incorrectly predicted in
previous simulations for high intensity operation of this device
are compromised.
\end{abstract}

%\begin{keyword}
%Heavy-ions; Gas stopping; Angular straggling; Gas collisions; Charge-exchange
%\PACS 24.10.Lx \sep 29.20.-c \sep 34.50.Bw \sep 29.25.Rm
%\end{keyword}
%\end{frontmatter}

% main text

\section{Introduction}
\label{intro}

The thermalization of energetic beams produced by fragmentation,
in-flight fission or fusion-evaporation reactions is an important
technology for the production of low-energy rare isotope beams. It
can provide access to radioactive beams of refractory or reactive
elements not amenable to the ISOL technique. These thermalizing
techniques are currently being aggressively developed and are
essential to the next generation of radioactive beam experiments.
The production of such beams would allow reaccelerated beam
experiments to venture into a regime of exotic isotopes beyond the
reach of current target/ion source assemblies, allowing for a
wealth of new experiments to be developed.

Thermalizing energetic ions in a gas volume and extracting them
for post-acceleration is a difficult task, but one that has now
been realized in many facilities interested in ISOL type
experiments \cite{ex1, ex2, ex3, ex4}. Such facilities typically cool
beams through energy loss in high-purity helium, guide the
thermalized ions to a nozzle where they are carried by gas flow
through a differential pumping aperture after which the heavy ions
are extracted and transported to a post-accelerator or used in
experiments involving cooled rare isotopes. While this approach
has been shown to work well at low intensity, most approaches have
shown limitations with incident beam intensity many orders of
magnitude lower than what is expected for next generation
radioactive beam facilities \cite{Limits1, Limits2}. The cyclotron stopper has been
suggested as a way to overcome this limitation.

\section{Proposed Concept}

The cyclotron stopper technique was proposed by Katayama et al.
\cite{Katayama} and a detailed study by Bollen et al.
\cite{Bollen} was recently published. This approach involves the
use of a weakly focusing cyclotron magnet which provides ions an
essentially unlimited path length due to cyclotron motion,
allowing the use of gas pressures significantly lower than current
techniques. As the reaction products ionize the helium gas they
slow down and spiral in towards the center where they eventually
reach thermal energies. The desired end effect is to concentrate
cooled heavy ions near the center of the magnet, away from the
bulk of the ionized helium.

The separation of cooled heavy ions
from the majority of ionized gas allows for efficient extraction
and removes several difficulties that have plagued most other gas
catcher schemes. The undesirable bulk of helium ions can then be
removed with the use of charge collection electrodes without
affecting the cooled fragments. Near the center of the magnet, the
thermalized heavy ions are carried towards an RFQ ion guide
through the use of a DC electric field, an RF-carpet, and gas flow.

\begin{figure}
\includegraphics*[scale = .5]{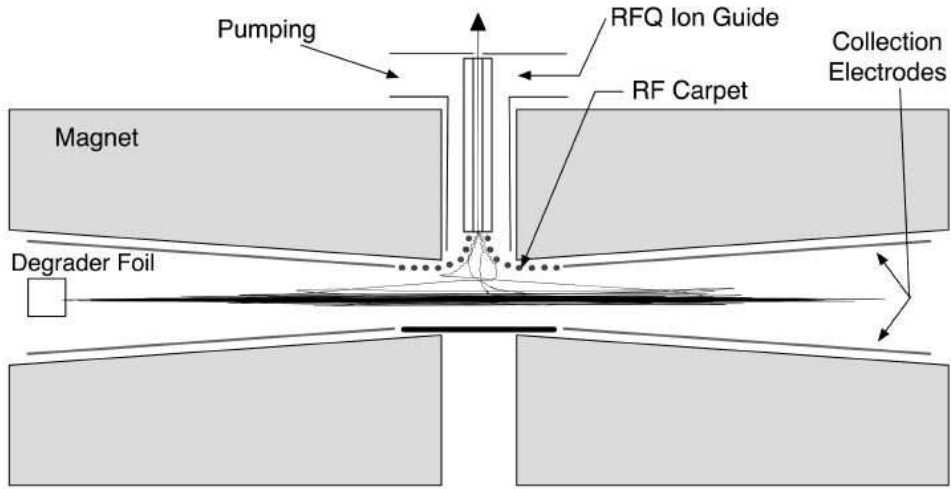}
\caption{Concept for the cyclotron gas-stopper studied in Ref. \cite{Bollen}.}
\label{cyclotron_fig}
\end{figure}

The main features that make such a concept attractive for use in
stopping energetic ions are its extremely long stopping length,
low gas pressure, and the clear separation between ionized helium
and cooled heavy ions. A study of this concept was reported in
Ref. \cite{Bollen} and has been used as a starting point for this
study. Fig. \ref{cyclotron_fig} illustrates a rough layout for the
concept proposed in Ref. \cite{Bollen}. Bollen et al. conclude that
for a prototypical Bromine radioactive beam a stopping efficiency
approaching 100\% can be reached in such a device with a 4 cm
gap between the charge collection electrodes. This result is surprising in that the lateral straggling of
such a beam coming to rest in gas, in the absence of magnetic or electric fields, far exceeds this dimension.

We therefore repeated the simulations while including various effects
that were either ignored or improperly treated in Ref.
\cite{Bollen}. In addition, we also performed simulations for
injection of beam with phase-space more representative of modern
large acceptance spectrometers. For the geometry and beam
properties used in Ref. \cite{Bollen}, we find a stopping
efficiency down by over two orders of magnitude compared to their
results. Our simulations indicate that a pole gap up to 18 times
larger is required to obtain sizable efficiencies. This not only
increases the technical difficulties with the magnet but also
removes some of the advantages that were proposed for operation at high
intensity. By modifying the injection scheme we can, in addition, increase the acceptance of the device by roughly an order of magnitude.

The approach taken below was to first reproduce the results of
Ref. \cite{Bollen}, then systematically improve on this simulation
by:
\begin{itemize}
\item Correcting the charge-changing collision method used in
that paper.
\item Including proper charge state of ions in solids for beam exiting the degrader.
\item Adding angular straggling.
\item Performing simulations with larger phase-space for the incoming
beam.
\item Investigating alternative means of focusing.
\end{itemize}
The basis for
the changes between the present simulations and the work of Ref.
\cite{Bollen} is given when each change is made and conclusions are stated on
the status of this approach as an alternative for high intensity
operation for both the original and the new larger acceptance geometry proposed here.

\section{Simulation}

\subsection{Basis For Simulation}

All of the simulation details discussed in this subsection, unless otherwise stated, have
been taken directly from Ref. \cite{Bollen} and have been used as
a starting point for simulations discussed later in this paper.
Information and details beyond what are discussed here can be
found in Ref \cite{Bollen}. The simulations are performed using a
Monte Carlo ray tracing code and consist of a weakly focusing
cyclotron magnet, 1 m in radius, filled with 10 mbar helium gas at zero degrees Celsius. The peak magnetic field of the magnet is $B_{o}$ = 2 T with a field index of $n_{o}$ = 0.2 and a radius of injection $r_{\scriptsize{\textrm{inj}}}$ = 0.8 m.

Bromine is used as a test case for
the simulation because of its moderate mass and because of the
charge-exchange data available for this element. The test beam
consists of 100 MeV/u $^{78}$Br passing through an aluminum
degrader resulting in a beam with an average energy of 610 MeV.
The initial simulation assumes that the beam exiting the degrader has an
average energy spread $\Delta$E/E of 20$\%$, a beam half width of
5 mm and a beam half divergence of 10 mrad.

The motion is modeled for non-relativistic ions using an axially
symmetric form $ \vec{B} ( \vec{ \rho } , z) = B_{\rho} \cdot
\hat{ \rho } + B_{z} \cdot \hat{ z }$ . Where, in a paraxial
approximation, $B_{\rho}$ and $B_{z}$ are given by:

\begin{equation}
B_{\rho} = -(n_{o} \cdot B_{o} / r_{\scriptsize{\textrm{inj}}})z
\label{eqno1}
\end{equation}

\begin{equation}
B_{z} = B_{o} - (n_{o} \cdot B_{o} / r_{\scriptsize{\textrm{inj}}})\rho
\label{eqno2}
\end{equation}

Energy loss is modeled using stopping tables for bromine ions in
helium provided by SRIM \cite{SRIM}  for a range of energies from
1 GeV to 100 eV. Energy straggling is not considered since it
corresponds to a deviation in the total path length of less than
5$\%$ and therefore does not have a sizable impact on the
simulation. Angular straggling was also not considered in the
original simulation because its effects were thought by \cite{Bollen} to be small. Verifying this assumption was one of the
leading motivations for trying to reproduce the simulations
reported in Ref. \cite{Bollen}. As will be evident later, the
effects due to angular straggling are not negligible.

Charge-exchange collisions of Br with helium are modeled for
single-electron exchange. Multiple-electron loss and capture can
be ignored since their cross-sections are in general orders of
magnitudes smaller. The average charge state $\bar{q}$ as a
function of ion velocity $v$ is calculated according to
semi-empirical formulae fitted to experimental data \cite{Betz},
\begin{equation}
\bar{q} / Z = \frac{\lg{(v / m_{1}Z^{\alpha_1})} }{ \lg{(n_{1}Z^{\alpha_2})}}   \; \; \; \;    (0.3 \leq \bar{q} / Z \leq 0.9)
\label{eqno3}
\end{equation}

\begin{equation}
\bar{q} / Z = AvZ^{-1/2}        \; \; \; \;  \; \; \; \; \; \; \;  \; \; \; \;   ( \bar{q} / Z < 0.3)
\label{eqno4}
\end{equation}
where $Z$ is the proton number and $m_{1}$, $n_{1}$, $\alpha_1$,
$\alpha_2$, and $A$ are parameters fitted to experimental data.

The cross-sections for charge loss $\sigma_{l}$ and capture
$\sigma_{c}$ for a given charge state $q$ are given by:
\begin{equation}
\sigma_{c} = \frac{\sigma_{o}}{2} \cdot e^{a_c(q - \bar{q})}
\label{eqno5}
\end{equation}
\begin{equation}
\sigma_{l} = \frac{\sigma_{o}}{2}  \cdot e^{a_l(\bar{q} - q)}
\label{eqno6}
\end{equation}
$\sigma_{o}$, $a_c$, and $a_l$ are parameters whose energy dependence have been parameterized by fitting Eqns. \ref{eqno5} $\&$ \ref{eqno6} to experimental data for Br in helium at a range of energies up to 14 MeV obtained from a review article by Betz et al. \cite{Betz}. $\sigma_{o}$ strongly determines the total cross section $\sigma$ for a charge exchange collision given by $\sigma = \sigma_c + \sigma_l$. Similarly, $a_l$ and $a_c$ determine the width of the charge state distribution for a given $\bar{q}$. The ratio of $\sigma_{c}$ and $\sigma_{l}$ to $\sigma$ give the probability for charge capture and charge loss, respectively, during a charge-exchange collision.

Beyond 14 MeV the values for $a_l$ and $a_c$ are fixed. This is not of much concern since the width of the charge distribution does not vary greatly
over a broad range of energies and, in addition, small variations
in the width of the charge distribution do not have a significant
effect on simulations. However, it is necessary to know
$\sigma_{o}$ or more importantly $\sigma$ for energies greater
than 14 MeV, since $\sigma$ directly determines the mean free path
$\lambda = 1/ ( n \cdot \sigma)$, where n is the number density of
the gas. If $\lambda$ is too large relative to the radius of
cyclotron motion, it will result in dramatic changes in trajectory
as an ion moves from one charge state to another. If instead
lambda is relatively small, the ions trajectory will be "smoothed
out" over the course of several charge exchange collisions,
resulting in an overall trajectory corresponding to the average
charge state $\bar{q}$.

The magnitude of $\sigma_{o}$ is determined for energies above 14 MeV and up to 700 MeV using Bohr's expressions  $\sigma_{c\_ \scriptsize{\textrm{Bohr}}}$ and $\sigma_{l\_ \scriptsize{\textrm{Bohr}}}$ for single-electron capture and loss by heavy ions in light targets given by Eqns. 4.5 $\&$ 4.6 of Ref. \cite{Betz}. 
\begin{equation}
\sigma_{c\_ \scriptsize{\textrm{Bohr}}} \sim  4\pi a_o^2 Z^{1/3}Z_T^5( v_o/ v )^6
\label{eqno8}
\end{equation}
\begin{equation}
\sigma_{l\_ \scriptsize{\textrm{Bohr}}} \sim  4\pi a_o^2 Z^{1/3}Z_T^2( v_o/ v )^3
\label{eqno7}
\end{equation}
Here, $v$ is the ion velocity, $a_o$ is the Bohr radius $a_o =
5.292 \cdot 10^{-9}$ cm, $v_o$ is the Bohr velocity $v_o = 2.118 \cdot 10^{8}$ cm/s, $Z$
is the atomic number of the projectile nucleus, and $Z_T$ is the
atomic number of the gas target. $\sigma_{o}$, determined in this way, combined with $\bar{q}$, $a_l$ and $a_c$, calculated as previously stated, are used in Eqns. \ref{eqno5} $\&$ \ref{eqno6} to give the values for $\sigma_{c}$ and $\sigma_{l}$ that are used to determine the probabilities for charge exchange in the simulation. This is different from what was
stated in Ref. \cite{Bollen}, where they claimed to use Eq. 4.7 of
Ref. \cite{Betz}.

\begin{equation}
\sigma_{l\_ \scriptsize{\textrm{Bohr}}} \sim \sigma_{c\_ \scriptsize{\textrm{Bohr}}} \sim \pi a_o^2 (Z^{1/3} + Z_T^{1/3})( v_o/ v )^2
\label{eqno9}
\end{equation}

Contrary to what is stated in Ref. \cite{Bollen}, Eq. \ref{eqno9}
is, as stated in Ref. \cite{Betz}, an equation for heavy ions in
heavy targets. For 610 MeV bromine the use of Eq. \ref{eqno9} in
calculating $\sigma$ results in a nearly 3 fold increase as
compared to calculations using Eqns.  \ref{eqno7} $\&$
\ref{eqno8}. In addition, it is reported in Ref. \cite{Bollen}
that 610 MeV bromine in 10 mbar of helium has a mean free path of
6 mm. However, 10 mbar of 273 K helium ($n = 2.6 \cdot
10^{17}$ atoms/cm$^3$) with a mean free path of about 6 mm corresponds
to a cross-section for charge-exchange of nearly $\sigma = 7 \cdot
10^{-18}$ cm$^2$. This is roughly 3 times larger than the value of
$\sigma$ calculated using Eq. \ref{eqno9} and about 9 times the
value of $\sigma$ given by the more appropriate Eqns. \ref{eqno7}
$\&$ \ref{eqno8}. We found no way of reconciling these
differences.

Furthermore, experimental values for charge-exchange
cross-sections provide an upper limit on reasonably acceptable
values. The charge-exchange cross-section for the mean charge
state of 8.5 MeV/u Kr$^{+27}$ in N$_2$ is measured to be $\sigma
\approx 4 \cdot 10^{-18}$ cm$^2$ \cite{data1}. In helium this
cross-section would be lower by a factor of two in order to
account for the diatomic nature of N$_2$ and an additional factor
of at least $(Z_{\scriptsize{\textrm{N}}} / Z_{\scriptsize{\textrm{He}}})^{2} \approx 10$ in accordance with Eqns. \ref{eqno7} $\&$ \ref{eqno8}. A similar measurement \cite{data2} reports a value of $\sigma \approx 2 \cdot 10^{-20}$
cm$^2$ for 8.55 MeV/u V$^{+21}$ in He. Both of these measurements
support the smaller values of $\sigma$ calculated using Eqns.
\ref{eqno7} $\&$ \ref{eqno8} (actually, this data points to even smaller
values of $\sigma$ than what is predicted by Eqns. \ref{eqno7} $\&$ \ref{eqno8} at these energies).

\subsection{Reproducing Previous Simulation}

\begin{figure}
\includegraphics*[scale = .4]{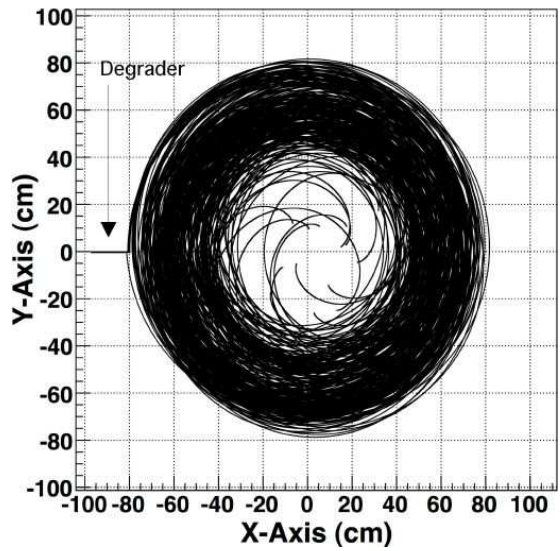}
\caption{Trajectories of 15 ions projected onto the x-y plane with
$\lambda$ limited to values less than 6 mm as used in Ref.
\cite{Bollen}. The degrader foil is located on the x-axis at x $\leq -79.5$ cm.} \label{simulation_1}
\end{figure}

\begin{figure}
\includegraphics*[scale = .4]{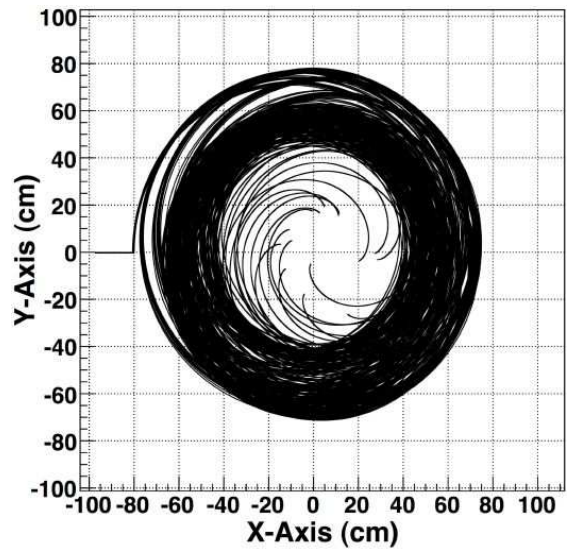}
\caption{Trajectories of 15 ions projected onto the x-y plane with
$\lambda$ limited to values less than 0.2 mm. The degrader foil is located on the x-axis at x $\leq -79.5$ cm.}
\label{simulation_2}
\end{figure}

Simulations have been produced exactly as described above
using SIMION ray tracing software \cite{SIMION} in an attempt to reproduce the results reported by Ref. \cite{Bollen} without any angular straggling from ion-gas
collisions. In order to match the parameters used by Ref. \cite{Bollen}, $\lambda$ was limited to values less than 6 mm, a value substantially smaller than those suggested by Eqns. \ref{eqno7} $\&$ \ref{eqno8} or data from experimental charge-exchange cross-sections. While the results of these simulations appear similar to those reported in Ref.\cite{Bollen}, the orbits are less regular, a small percentage of ions are lost due to collisions with the degrader as they cycle through the first few cyclotron orbits, and the final ion position distribution does not form a distinct toroidal stopping volume. This is due to the fact that the trajectories of the ions are not completely "smoothed" out from charge-exchange collisions, which is clearly visible when the trajectories from this simulation presented in Fig. \ref{simulation_1} are compared directly to Fig. \ref{simulation_2} where, to better reproduce the results of their simulations, $\lambda$ was arbitrarily further limited to values below 0.2 mm (corresponding to $\sigma_o > 2 \cdot 10^{-16}$ cm$^2$).

The trajectories of Fig. \ref{simulation_2} are in perfect agreement with Ref. \cite{Bollen}, where the only appreciable differences in ion trajectories are due to the momentum and angular spread of ions exiting the degrader. Furthermore, the maximum density of deposited energy per ion is found to be about 125 keV/cm$^2$ with a clear separation between the bulk of ionization and the 30 cm radius toroidal stopping volume with a 3 cm height, again, this is in
perfect agreement with Ref. \cite{Bollen}. As previously predicted, such a configuration should be able, if these simulations were correct, to quickly remove the bulk of ionized helium ions across the 4 cm electrode gap fast enough to support beam rates as high as 10$^8$/s.

\subsection{Necessary Improvements}

The simulations discussed in the previous subsection have not
addressed several unavoidable problems. Thus far, the mean free
path for charge-exchange collisions has been artificially limited
to small values. It is necessary to use cross-sections that agree
with experimental results as well as theory, such as
cross-sections calculated using Eqns. \ref{eqno7} $\&$
\ref{eqno8}. In addition, the path length between charge-exchange
collisions is assumed to be the same as the mean free path. This
is not realistic and fails to accommodate the fact that the path
length should vary about the mean free path. A more appropriate
method is to sample the path length $l$ using the following
formula:
\begin{equation}
l = - \lambda \cdot ln \; p
\label{eqno10}
\end{equation}
Where $\lambda$ is the mean free path between charge-exchange
collisions discussed earlier and $p$ is a random number between 0
and 1.

Fig. \ref{simulation_1} demonstrates that in the regime where $\lambda$ is larger than 6 mm
the effects on the ion trajectories due to charge-exchange
collisions are not "smoothed out". It is necessary, therefore, to
recognize that ions exiting a solid degrader will have a mean
charge state higher than the mean charge state for the same ions
in a gas. This will have an appreciable effect if ions travel a
sizable fraction of an orbit before returning to the mean charge
state. The average charge state for ions in a solid can be
calculated using Eq. \ref{eqno3}  with $m_{1}$, $n_{1}$,
$\alpha_1$ and $\alpha_2$ set to the appropriate values listed in
Ref. \cite{Betz} for bromine in solids. Typically the relative
charge state is greater by 5$\%$ and sometimes as high as 100$\%$
above that of a gas.

Another point of concern is that angular straggling due to
scattering in helium has been disregarded. Simulations in TRIM \cite{SRIM} predict that 610 MeV bromine ions coming to rest in 10 mbar of helium at 273 K scatter on average about 37 cm
radially with a standard deviation of 20 cm over the course of their roughly 65 m stopping length. In
the previous simulation it was demonstrated that a beam half
divergence of 10 mrad results in an extraction region with a 3 cm
height. It seems reasonable, therefore, to expect that angular
straggling would add significantly to this. The real scenario is
obviously more complicated as the magnet provides a source of weak focusing which would diminish this effect somewhat. Although, this focusing becomes much weaker as the ion's charge state decreases while they lose energy. In addition, as the ions fall inward towards the end of their trajectory their motion becomes mostly radial, which is not focused vertically at all. As most angular straggling occurs towards the end of the trajectory, it is difficult to
imagine that angular straggling due to ion-gas collisions would
have a negligible effect.

In order to produce realistic angular straggling modeling, we have based our simulations on data from TRIM, which is itself based on an extensive experimental data set. TRIM yields, with no adjustable parameters, trajectories for arbitrary ions stopping in solids or gases. TRIM agrees well with experimental data for stopping ranges in solids and light gasses and scattering in solids, but there is very little data available for comparison with scattering in light gases. However, we have used TRIM together with the procedure presented below to reproduce experimental data for heavy ions in gas-filled magnet applications at low energies \cite{Paul,Rehm} and find very good agreement. In particular, the effects directly related to angular straggling, which can be isolated in some applications, are satisfactorily reproduced. Additional data in this region would be extremely valuable in determining both the uncertainty in TRIM calculations and in further validating the results of this work.

TRIM  allows one to produce detailed collision data files that include the position and energy of the projectile ion for each collision that results in the displacement of a target atom. Needless to say, there are many such collisions. This collision file can be used to produce normalized velocity
vectors at the vertex of each collision and by comparing the vectors
at each vertex to the subsequent vector, it is possible to
calculate the scattering angle at each vertex.

It is necessary to recognize, however, that the data produced by this method contains artifacts used to hasten the calculation, which convolute the data below certain energies and results in low precision data for interactions on length scales small compared to the viewing window. One can overcome this by running TRIM calculations at multiple energies, narrowing the viewing window as the energies are stepped down, and then piecing together the good high resolution data. The high resolution data can then be used to fit the collision distances and scattering angles as functions of the ion's energy.

\begin{figure}
\includegraphics*[scale = .4]{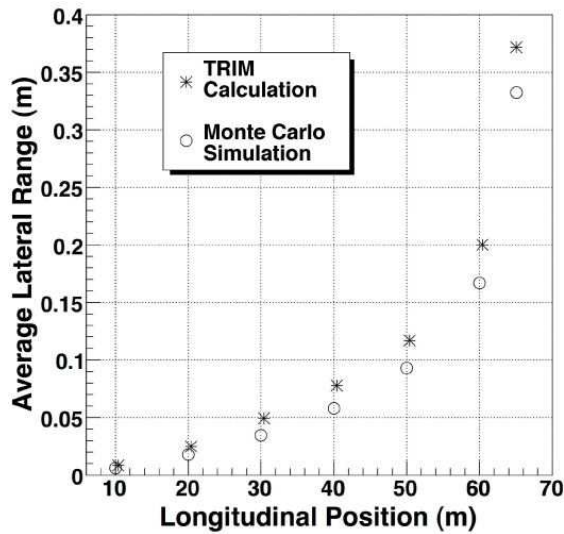}
\caption{Plot of the average lateral range as a function of the ions longitudinal position for the Monte Carlo simulation run in SIMION and the original TRIM calculations.}
\label{TRIM_CALC}
\end{figure}

Monte Carlo simulations using the method described above, incorporated into SIMION, produce ion trajectories
that agree very well with TRIM calculations. In the absence of a magnetic field, over the 65 m stopping length, 610 MeV bromine ions scattered on average about 33 cm transversely with a standard deviation
of about 18 cm, similar to the results of the TRIM calculations stated earlier. The results of this Monte Carlo test simulation are illustrated in Fig. \ref{TRIM_CALC}, which shows the average lateral range as a function of the ions longitudinal position compared to calculations performed by TRIM.

\subsection{Improved Simulations}

\begin{figure}
\includegraphics*[scale = .4]{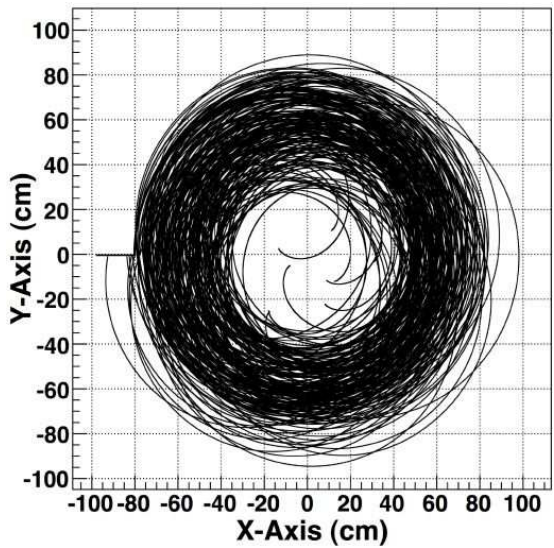}
\caption{Trajectories of 15 ions projected onto the x-y plane
using values for $\sigma$ calculated according to Eqns.
\ref{eqno7} $\&$ \ref{eqno8}.}
\label{simulation_3}
\end{figure}

\begin{figure}
\includegraphics*[scale = .4]{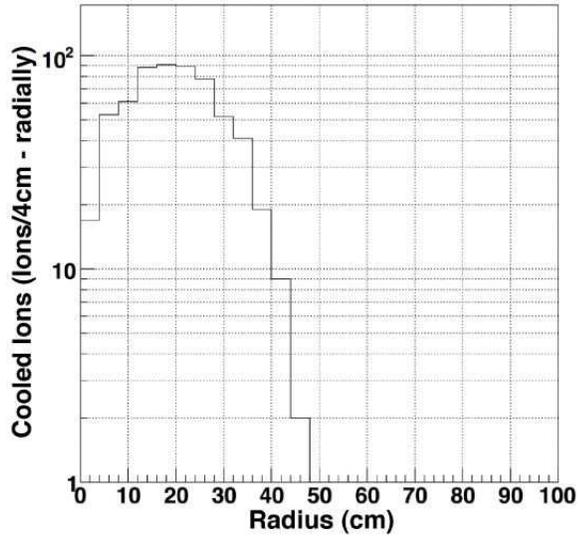}
\caption{Histogram of the 600 cooled ions along radial direction. The
density is plotted in units of ions/(4 cm - radially).}
\label{ion_density_3}
\end{figure}

\begin{figure}
\includegraphics*[scale = .4]{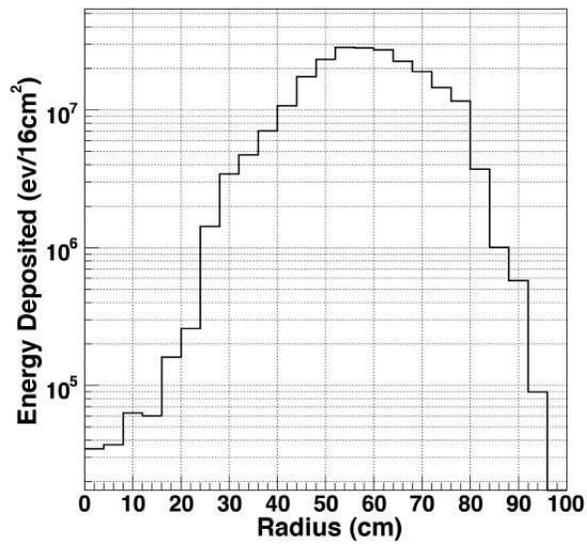}
\caption{Total energy deposition radially along x-y plane for 50 bromine
ions of which 20 hit the degrader, plotted on a
logarithmic scale in units of eV/(16 cm$^2$).}
\label{energy_dep_3}
\end{figure}

Let us first resolve the issue of charge-exchange cross-sections
by using values for $\sigma$ calculated according to Eqns.
\ref{eqno7} $\&$ \ref{eqno8} at energies greater than 14 MeV. The
results of this simulation are shown in Figs. \ref{simulation_3},
\ref{ion_density_3} $\&$ \ref{energy_dep_3}. In order to make the
simulation more robust, straggling in the charge-exchange path
length has been introduced according to the method discussed in
the previous subsection. This addition has only a small effect on
the results of the simulations.

As expected, these trajectories are much rougher than those in the previously demonstrated simulations. The spread in effective magnetic
rigidities due to ion-gas collisions agrees very well with
experimental measurements performed with the use of a gas-filled
Enge spectrograph such as those reported by Paul et al.
\cite{Paul}. In contrast,  we have not been able to reconcile the previously reported "smoothed out"
trajectories with either theory or
experimental measurements.

In our simulation, 40$\%$ of the ions hit the degrader,
significantly reducing the efficiency. In addition, the variation
in trajectories due to the increased path length between
charge-exchange collisions causes a slight broadening of the
extraction region as well as an increase in the energy deposition
within the extraction region. The ions are cooled within a disc,
35 cm in radius and 3 cm in height. Fig. \ref{energy_dep_3} shows
the energy deposited for a simulation where 20 of the 50 ions hit
the degrader. In this simulation 407 MeV was deposited within the
extraction region. Thus, there are about 331,000 helium ions or 29
helium ions per cm$^3$ produced within the extraction region per
cooled bromine ion. The peak density of deposited energy is less than what was investigated in the previous simulation and is therefore not of concern.

In subsequent simulations the ions were initialized with a mean
charge state for bromine in solids. This was found to have little
effect. In a run consisting of 1000 ions, 403 hit the degrader, a
negligible change from the previous simulation. This is not too
surprising when one considers that the cross-section for
charge-exchange collisions is dominated by an exponential as ions
drifts away from the mean charge state.

\begin{figure}
\includegraphics*[scale = .45]{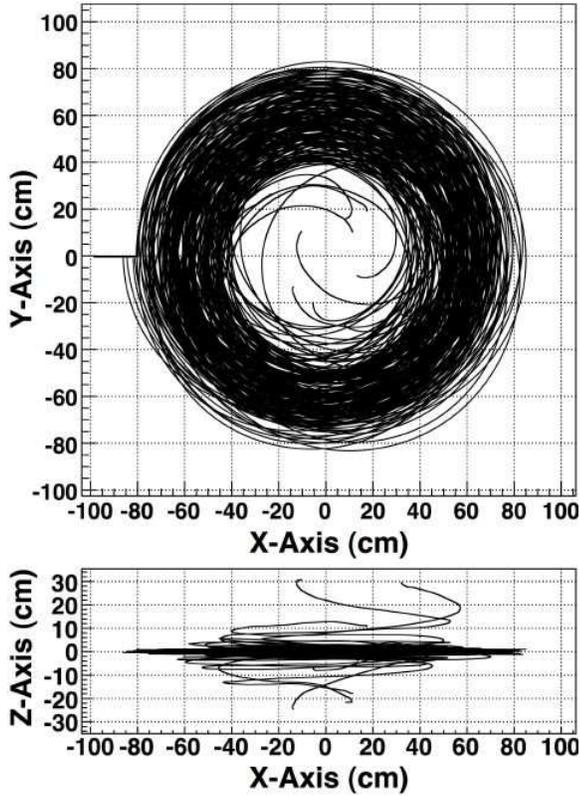}
\caption{Trajectories of 15 ions with angular straggling included projected onto the x-y $\&$ x-z planes.}
\label{simulation_5}
\end{figure}

\begin{figure}
\includegraphics*[scale = .4]{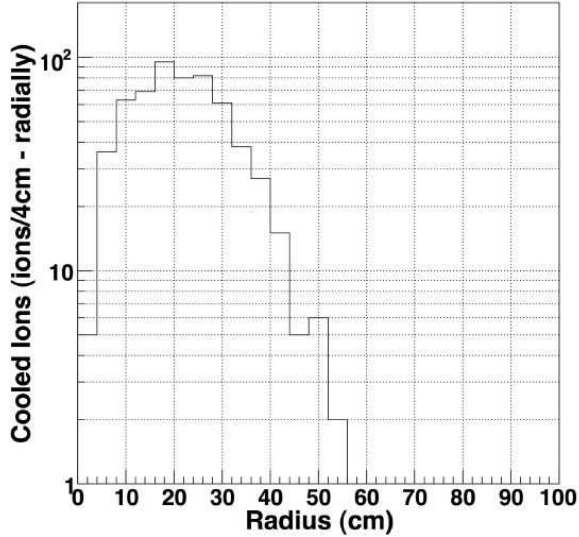}
\caption{Histogram of the 584 cooled ions along the radial direction plotted in units of ions/(16 cm$^2$).}
\label{ion_density_5}
\end{figure}

The most drastic change to the simulation comes from the addition
of angular straggling, introduced using the method described in
the previous subsection. Unlike the previous simulations, it is
important to note that the pole gap of the magnet used in the
simulation was increased to 70 cm as most of the ions were hitting
the top and bottom of the magnet. Fig. \ref{simulation_5} shows the trajectories of 15 ions
projected onto the x-y and x-z planes of which 6 ions hit the degrader. Fig. \ref{ion_density_5} illustrates a density plot of the cooled ions
for a run consisting of 1000 ions, in which 416 hit the degrader
or the walls of the magnet.

\begin{figure}
\includegraphics*[scale = .4]{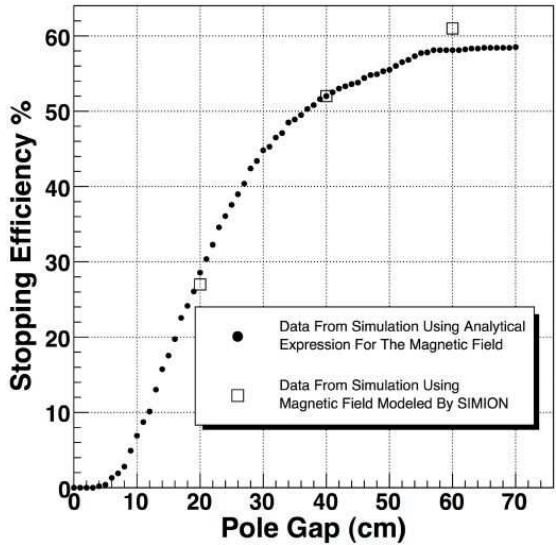}
\caption{Stopping efficiency as a function of magnet pole gap.}
\label{gap_width_5}
\end{figure}

In these simulations the magnetic field was calculated using Eqns.
\ref{eqno1} $\&$ \ref{eqno2}, which were derived in the paraxial
approximation. It is not clear that these expressions are still
valid for simulations of such large gap magnets. In order to
verify their validity, the magnetic fields of three large gap
magnets of the same specifications and 1 m radii, but varying pole gaps, were modeled in SIMION. Fig.
\ref{gap_width_5} shows the percentage of cooled ions as a
function of the magnet pole gap for simulations using both the
analytical expressions of Eqns. \ref{eqno1} $\&$ \ref{eqno2} and
the magnetic fields produced by SIMION. There are only a few data
points for simulations using the SIMION modeled magnetic fields
since these simulations require a new magnet geometry to be
compiled for each specific pole gap. Fig. \ref{gap_width_5}
demonstrates that the analytical expressions used to calculate the
magnetic fields are sufficient even for the largest pole gaps
used in these simulations.

From Fig. \ref{gap_width_5}, it is clear that a stopping efficiency below 1$\%$ is obtained with the previously proposed 4 cm electrode gap. The maximum achievable stopping efficiency
of $60\%$ is limited mainly by losses in the degrader. This comes before taking into account further losses due to
neutralization and extraction. Achieving efficiencies in this range requires the construction of a 2 Tesla
cyclotron magnet with a 60 cm pole gap. In practice, the actual gap would have to be of the order of 70 cm in order to house the necessary equipment that such a concept would require. This poses
many difficult engineering challenges as well as result in a significant
increase in cost.

\begin{figure}
\includegraphics*[scale = .45]{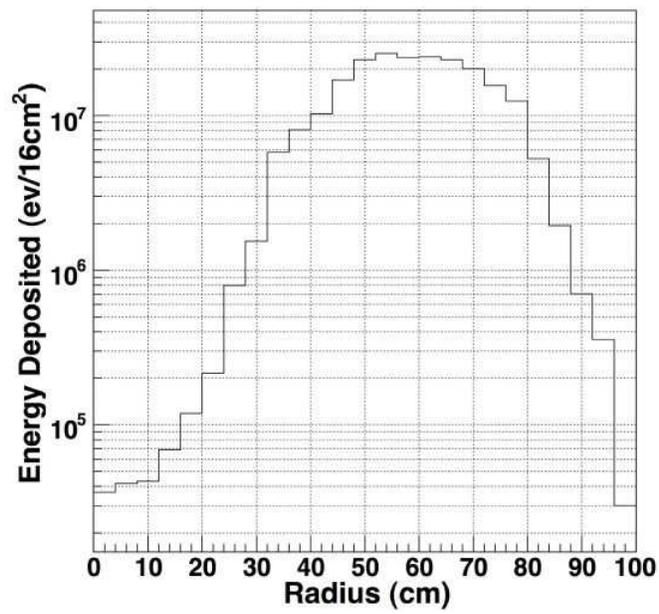}
\caption{Total energy deposition along radial direction for 50 bromine
ions of which only 27 were successfully cooled. The density is
plotted on a logarithmic scale in units of eV/(16 cm$^2$).}
\label{energy_dep_5}
\end{figure}

Fig. \ref{energy_dep_5}  shows the
energy deposited from a run in which 27 out of 50 ions were
successfully cooled. The maximum density of deposited energy per cooled ion is found to be about 70 keV/cm$^2$, a reduction by a factor of about 1.8 compared to the simulation illustrated in Fig. \ref{energy_dep_3}. In the previous simulations this charge was confined within the 4 cm gap, allowing for quick extraction of bulk Helium ions. A new analysis is required, however, for this new magnet geometry. According to
 \cite{Bollen} the He ion space-charge, while moving towards the charge collection electrodes, induces a voltage:
\begin{equation}
V_{\scriptsize{\textrm{ind}}} = \sqrt { \frac{eQ}{4\varepsilon_{0} k} } d^{2}
\label{eqno11}
\end{equation}
Where, $e$ is the electric charge, $Q$ is the rate of ionization per volume, $\varepsilon_{0}$ is the permittivity of free space, $d$ is the distance between charge collection electrodes, and $k = k_{0}(\textrm{He}) \cdot (1013$ mbar$)/P_{\scriptsize{\textrm{He}}}$ is the mobility of helium ions at a given pressure $P_{\scriptsize{\textrm{He}}}$ ($k_{0}(\textrm{He}) = 1.05 \cdot 10^{-3}$ m$^{2}$/(Vs) ).

For a 60 cm gap with helium pressure of
10 mbar, the Paschen curve in helium indicates a breakdown voltage
of roughly 1800 volts, probably lower in the presence of magnetic
fields. This corresponds to an electric field of about 30 V/cm across
the gap, the same electric field used in the previous rate calculations. Thus, we find that while the reduction in the peak density of deposited energy will increase the beam intensity tolerance by a factor of roughly 1.8, the
15 times larger gap will reduce the ion intensity this
device can tolerate by about a factor of 15. An overall reduction of about a factor of 10 is expected within the formalism of Eqn. \ref{eqno11}. However, with the gap size
being now larger than the radial distance between the stopping
volume and the bulk of the ionization, it is to be expected that
some of the ionization will be repelled into the extraction volume. While this can easily be
handled if the ionization remains in the form of helium ions since those are not transported by the
RF carpet, even a small fraction of this ionization creating contaminant ions by charge exchange 
processes could easily saturate the extraction region. A more complete description and accurate 
calculation of this effect is presented later in the text for a proposed
improved geometry with more details given in
\cite{sternberginprep} in preparation.

Neglecting space-charge effects, extraction times for Br ions become only slightly longer across the 70 cm gap. For a 30 V/cm electric field the ion drift velocity is about 600 m/s. Using this
result together with the times for extraction along the RF carpet,
funnel, and linear RFQ guide to a point outside the magnet as
reported in Ref. \cite{Bollen}, one finds that it would take
roughly 5-10 ms to extract ions across the 70 cm pole gap, comparable to the times stated in Ref. \cite{Bollen}.

These calculations do not take into account effects on extraction times
that arise from the build up of space charge within the extraction region. Within the extraction region, a disc 40 cm in radius and 70 cm in
height, a total of 1.25 GeV was deposited for the run in which 27 out of 50 ions were successfully cooled,
corresponding to about $1.1 \cdot 10^6$ helium ions per cooled bromine ion or about 3
helium ions per cm$^3$. This charge as well as charge flowing into the extraction region from the region of bulk ionization will certainly result in a reduction of extraction efficiencies, a reduction in allowed beam intensities, and an increase in extraction times.

All of the simulations discussed thus far have been performed
using a small beam-spot with an average energy spread $\Delta$E/E
of 20$\%$ at 8 MeV/u, a beam half width of 5 mm and a beam half
divergence of 10 mrad after passing through the degrader foil.
These parameters account for only a small fraction of
the desired reaction products which are produced with a large
variance in both beam energy and angular divergence. The fragment
separator in use at current medium energy fragmentation facilities have momentum acceptance of roughly 5\% with angular acceptances
of $\pm$ 50 mrad in both the x and y directions for a 1 mm spot. Various techniques for momentum compression have  been discussed at great length in the literature \cite{Geissel, Weick, Scheidenberger} and designs for proposed new facilities \cite{fragRIA} extend this momentum acceptance to 10$\%$ with similar angular acceptance. An initial beam energy of 150-200 MeV/u is required to yield the 100 MeV/u mass separated recoils at the end of the fragment separator. The selection
process through the fragment separator will have the recoils going
through degraders of roughly half the range of the selected
fragment so that according to \cite{Dufour} the beam-spot size at
the end of the separator is increased by a factor of 2 over that
without degraders. The initial 1 mm beam-spot with $\pm$ 50 mrad
angular envelope and no energy spread therefore fills already at this intermediate
energy the full 1 cm beam-spot with $\pm$ 10 mrad assumed in
\cite{Bollen} and used so far in this paper. Adding the effect of
the 5$\%$ momentum spread acceptance of current fragment separators, a momentum dispersive system and a wedged monochromatic
degrader are required to minimize the energy spread at the degrader. The small angular acceptance used in the simulations limits the resolution that can be achieved in this last stage so that a momentum dispersion of at least 2 cm/$\%$ is required leading to a beam-spot size of at least 10 cm high (dispersion is along the vertical axis) by 1 cm wide on the degrader, even for current separators.

\begin{figure}
\includegraphics*[scale = .4]{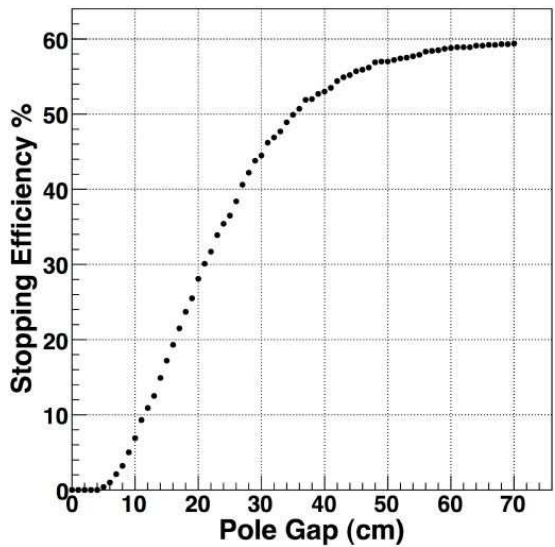}
\caption{Stopping efficiency as a function of pole gap for
simulation using a 1 cm by 10 cm beam-spot.}
 \label{gap_width_6}
\end{figure}

Fig. \ref{gap_width_6} shows the percentage of cooled ions as a
function of the magnet pole gap for the simulation performed with
the 1 cm by 10 cm beam. It could have been
chosen to widen the beam along the x-axis rather than the z-axis,
however, this would result in very limited gains as most of the
added beam would return and hit the degrader foil.

Contrary to what one would naively expect, Fig. \ref{gap_width_6} shows a
slight increase in the maximum stopping efficiency as compared to Fig. \ref{gap_width_5}. There are several
factors that contribute to this. First, the relative phase of
oscillations along the z-axis and x-axis are aligned such that the
maximum velocity along the z-direction, and therefore the minimum
in the x-y plane, is achieved after slightly more than half a
rotation in the x-y plane. Thus, as ions finish their first
rotation they do so with a slightly reduced cyclotron radius. This
reduces the likelihood of ions hitting the degrader, particularly
for ions that start further off-center. Furthermore, the
limitations in the efficiency of the system are dominated by the
loss of ions due to large angular straggling as well as losses due to ions hitting the degrader after one or more rotations. The restoring force of the magnet along the
z-direction is such that ions starting at $z = 0$ with 10 mrad
divergence relative to the x-y plane will oscillate back and forth
from $\pm$1.5 cm. When the ions start more than 2 cm off-center
along the z-axis the ions will oscillate with an amplitude
approximately equal to their starting position. It is only in the
final several meters of the ions flight that appreciable
scattering takes place resulting in the shape of the curves seen
in Figs. \ref{gap_width_5} $\&$ \ref{gap_width_6}.

\begin{figure}
\includegraphics*[scale = .4]{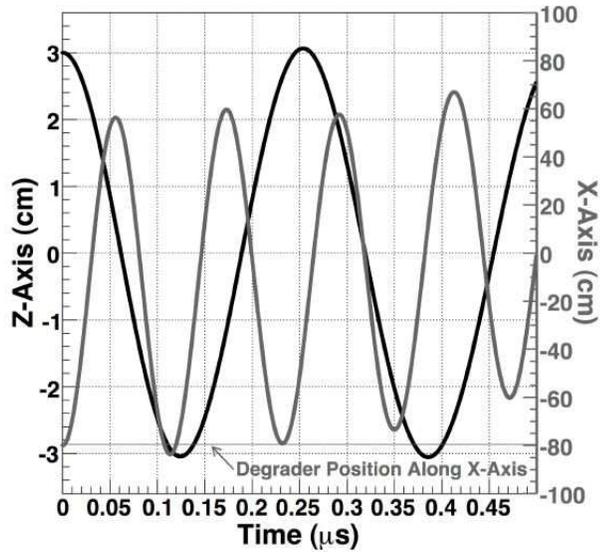}
\caption{Position along z-axis(black) and x-axis(grey) as a function of time. The grey line at the bottom represents the degrader located at x $\leq -79.5$ cm.}
\label{xzt_plot}
\end{figure}

A thorough analysis of the relative phase of oscillations along
the z and x directions reveals a method for limiting losses in the
degrader. Fig. \ref{xzt_plot} illustrates initial oscillations
along both axis for an ion starting at z = 3 cm. From this plot
it is evident that the phases of the two oscillations are aligned
such that an ion starting at z = 3 cm reaches z = -3 cm very close
to where the ion finishes its first rotation in the x-y plane.
That is to say, oscillations along the z-axis occur at about twice the
cyclotron period. In this type of configuration the ion
illustrated in Fig. \ref{xzt_plot} would hit the degrader as it
returned on its first rotation. However, if one moves the incoming
beam and the degrader so they are located at z $\geq$ 0, the same
ion misses the degrader on its first pass and the combination of
cooling and a slight procession in the orbit causes it to just
miss the degrader on the second pass. At injection the cyclotron frequency and the frequency of oscillations in the z-direction (in the paraxial and small angle approximations) are given by:
\begin{equation}
\omega_{c} \approx \frac{qB_{o}(1-n_{o})}{m}
\label{eqno12}
\end{equation}
\begin{equation}
\omega_{z} \approx \sqrt{ \frac{qvB_{o}n_{o}}{mr_{\scriptsize{\textrm{inj}}}}}
\label{eqno13}
\end{equation}
The ratio of the frequencies is given by:
\begin{equation}
\frac{\omega_{z}}{\omega_{c}} \approx \sqrt{\frac{mvn_{o}}{qB_{o}r_{\scriptsize{\textrm{inj}}}}}(1-n_{o})^{-1} = \sqrt{\frac{r_{c}n_{o}}{r_{\scriptsize{\textrm{inj}}}(1-n_{o})}}
\label{eqno14}
\end{equation}
Where $v$ is the initial ion velocity, $m$ is the ion's mass, $q$ is the charge of the ion, $n_{o}$ is the field index (as defined in Eqns. \ref{eqno1} $\&$ \ref{eqno2}), $B_{o}$ is the peak magnetic field, $r_{\scriptsize{\textrm{inj}}}$ is the injection radius, and $r_{c}$ is the characteristic cyclotron radius of the ion upon injection. Thus, the ratio of the vertical focusing frequency versus the cyclotron frequency for the initial orbits is a constant for this magnetic field configuration for all particles whose initial cyclotron radius corresponds to the .8 m injection radius used in this configuration. This approach of offsetting in the vertical direction the injection point and degrader is therefor generally applicable to all ion species by setting the injection energy, or scaling the magnetic field, to obtain the appropriate initial cyclotron radius.

\begin{figure}
\includegraphics*[scale = .4]{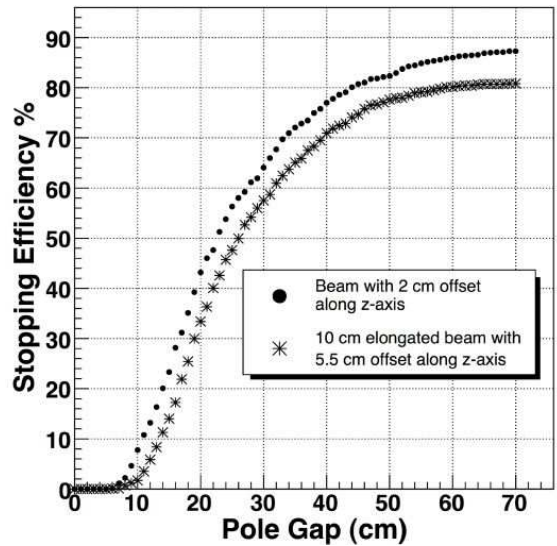}
\caption{Stopping efficiency as a function of pole gap using offset beams.}
\label{gap_width_7}
\end{figure}

Fig. \ref{gap_width_7}
illustrates the stopping efficiency for two simulations. The first
simulation has a 1 cm beam-spot offset by 2 cm along the z-axis,
with the degrader located at z $\geq$ 1.5 cm. The second
simulation has a beam-spot 1 cm wide, 10 cm tall, and is offset by
5.5 cm along the z-axis, with the degrader located at z $\geq$ 0.5
cm. These types of configurations significantly reduce losses in
the degrader and greatly improve the overall stopping efficiencies
(but the large magnet gap is still required). A maximum achievable stopping efficiency approaching 90$\%$ and 80$\%$ has been observed for the 2 cm offset beam and the much larger acceptance offset 1 cm by 10 cm beam, respectively, for a 70 cm gap. The cost and design complications of such a magnet would be high and require one to consider the tradeoff between efficiency and the size of the pole gap. A more modest 30 cm gap would allow for a stopping efficiency of about 65$\%$ and 60$\%$ for the 2 cm offset beam and the offset 1 cm by 10 cm beam respectively.

A detailed analysis of the space charge effect has been performed for a 60 cm pole gap with an offset large acceptance 1 cm by 10 cm beam. The energy deposited by the decelerating ions is converted to a charge creation rate per unit volume using the known average energy required per ion-electron pair creation of 42 eV in helium gas. The electrons move roughly 1000 times faster than the helium ions in the gas under the influence of an electric field and are neglected since they only contribute insignificantly to any space charge build-up. The trajectories of the He ions created by the energy loss are calculated in the electric field created by the electrodes (and the magnetic field) and a "static" space charge distribution is built from the time averaged space-charge density they create. The electric fields created by this space-charge distribution is then calculated. The procedure is then repeated with further ion creation and transport now in the combined electrode-generated and space-charge-generated electric fields and the resulting space-charge density is recalculated. The procedure is repeated until the problem converges to a self-consistent solution (more details will be available in \cite{sternberginprep}).

\begin{figure}
\includegraphics*[scale = .4]{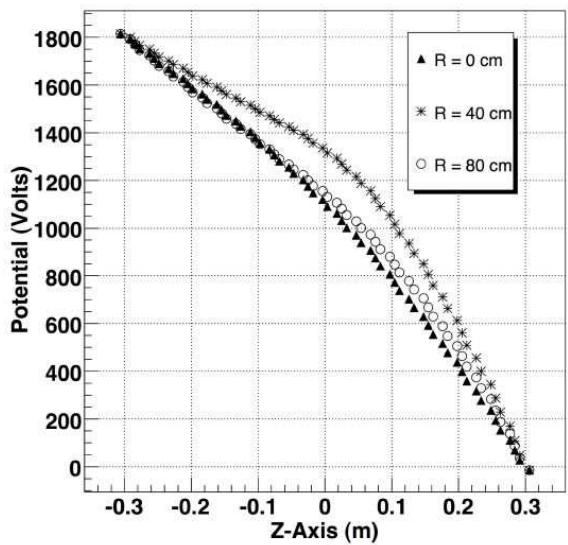}
\caption{Total electrostatic potential along z-axis coming from both the applied collection potential and the Induced space charge potential at various radial distances for $1 \cdot 10^{8}$ incident Br ions per second.}
\label{1e8_beam}
\end{figure}

\begin{figure}
\includegraphics*[scale = .4]{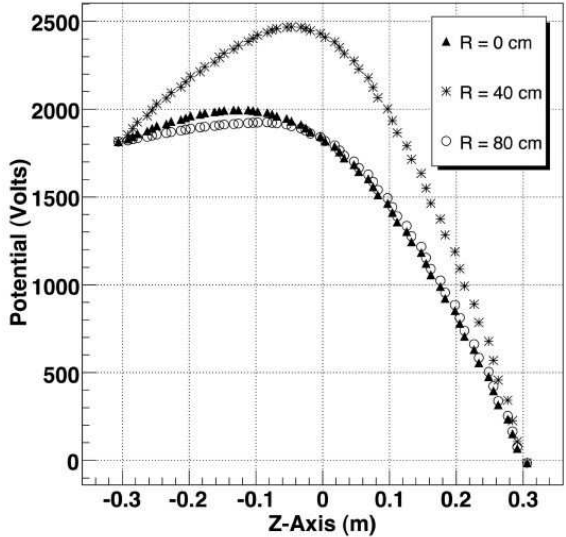}
\caption{Total electrostatic potential along z-axis coming from both the applied collection potential and the Induced space charge potential at various radial distances for $5 \cdot 10^{8}$ incident Br ions per second.}
\label{5e8_beam}
\end{figure}

The main results of these calculations for the geometry of interest are shown in Figs. \ref{1e8_beam} $\&$ \ref{5e8_beam}, which depict the applied DC potentials modified by space-charge buildup at different radii for incoming ion beam intensities of $1 \cdot  10^{8}$ and $5 \cdot  10^{8}$ ions per second respectively. At $1 \cdot  10^{8}$ ions per second, the fields are modified but essentially behave like the case with no space charge where the plot would show a straight line from 1800 to 0 volts going across the plot. All ions are pushed towards the extraction region and fields are large enough to avoid neutralization and keep extraction times short enough to limit charge exchange with contaminants.

In the extraction region the ions are transported along a 40 cm radius RF-carpet towards an extraction RFQ. The DC-field perpendicular to the RF-carpet is on the order of 40 V/cm, similar to the case where no space-charge is present. While no RF-carpet has operated under these exact conditions, the general scaling properties of the RF-force are well known \cite{ex2} and for the correct pressure and frequency regime the force goes as one over the pressure squared and is proportional to the ion mass. Operation with gradients in excess of 20 V/cm for masses in the range of roughly 100 AMU has been demonstrated at pressures in excess of 100 mbar with an RF-cone operating at 2 MHz \cite{Savard2}, as has operation at fields of up to 18 V/cm at a pressure of about 100 mbar for ions of mass 8 with an RF-carpet operating at 15 MHz \cite{ex2}. Both suggest, when taking into account the much lower 10 mbar operating pressure required here, the ability to handle fields well in excess of 50 V/cm for the mid mass nuclei considered here. It is expected from these results that the device can operate without severe disturbance at $1 \cdot  10^{8}$ incoming ions per second with little effect on extraction time and efficiency. 

At $5  \cdot  10^{8}$ ions/s, the potential from the space charge is severe enough to have created a potential maximum with the ions in the lower half of the device (the negative z-axis on the plot) now pushed down, away from the extraction RFQ, instead of up. In the upper half (the positive z-axis of the plot) where the ions are still directed towards the extraction RFQ, the electric gradient is now much larger, which may lead to a decrease in the Br extraction time. The RF-focusing structure on the pole may still be strong enough to repel the heavy ions because of the low operating gas pressure (the He ions are not repelled). More importantly, helium ions created in the central region where the electric potential has a maximum spend a large period of time, up to hundreds of millisecond, in the gas before being extracted and are much more likely to charge exchange with contaminants in the gas which will then be carried out to the extraction RFQ and lead to saturation. Radioactive ions stopped in the resulting low electric field central region are also more likely to neutralize or undergo charge-exchange reactions leading to further losses. All of these effects are still negligible at $1 \cdot 10^{8}$ incoming ions per second which we determine as the practical operation limit for this geometry. This device has a larger stopping range than any demonstrated gas catchers and a space-charge limit that far exceeds that demonstrated for all other large gas catcher except for the recent high-intensity gas catcher developed at ANL \cite{EMIS07}. The large stopping range make the proposed cyclotron stopper especially well suited to the lightest species who have low stopping power and are therefore difficult to stop in a linear gas catcher.

Alternative methods of focusing have also been investigated in an
attempt to minimize the effects of angular straggling. Simulations
using magnets with increased field indexes failed to produce
significant increases in efficiencies. However, split sector
magnets were found to have several benefits.

\begin{figure}
\includegraphics*[scale = .4]{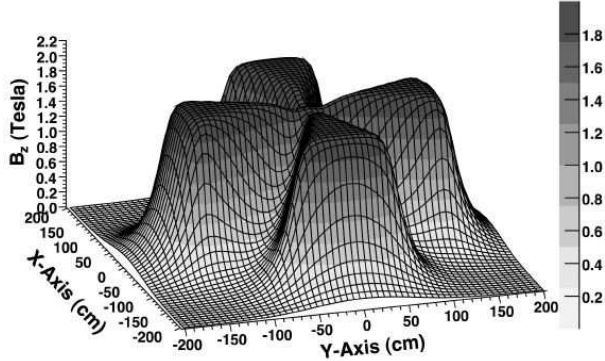}
\caption{Map of $B_{z}$ for a split sector magnet with a 30 cm gap
and 20 cm poles. The field strength is plotted on a linear
scale and reaches a maximum value of 2 Tesla.}
\label{Field_Map}
\end{figure}

Split sector magnets and their fields were modeled in SIMION. Fig
\ref{Field_Map} shows a mapping of $B_{z}$ along the primary plane
of a 1.5 m radius split sector magnet with a 30 cm pole gap, 20 cm wide pole
magnets, and a nominal field strength of 2 Tesla. The split sector
magnet simulations required a few adjustments from previous
simulations, namely, the optimal radius of injection was found to
be between 90 $\&$ 110 cm depending on the gap of the magnet.

\begin{figure}
\includegraphics*[scale = .45]{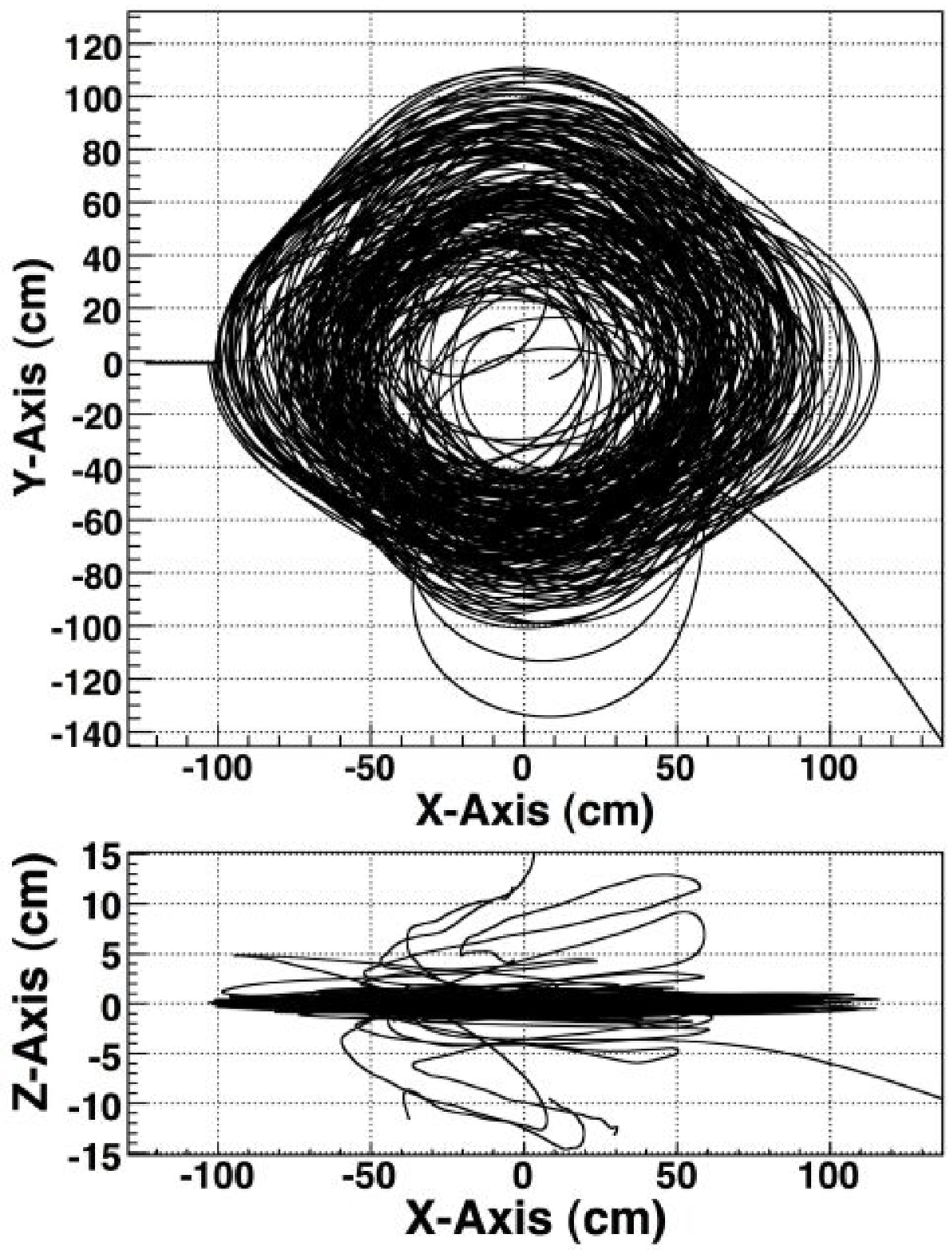}
\caption{Trajectories of 15 ions projected onto the x-y $\&$ x-z planes using split sector magnet with 30 cm pole gap.}
\label{simulation_7}
\end{figure}

\begin{figure}
\includegraphics*[scale = .4]{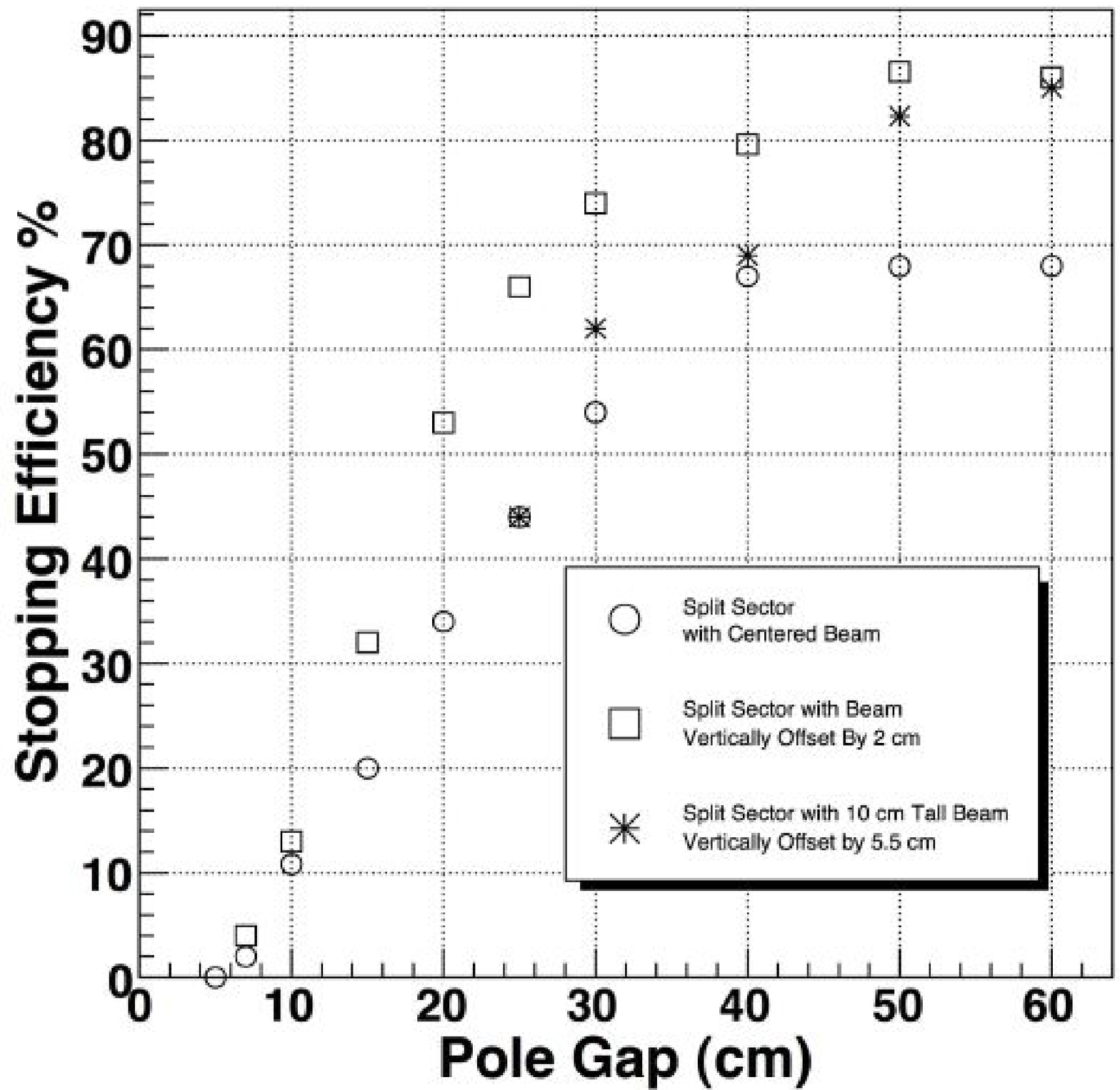}
\caption{Stopping efficiency as a function of pole gap.}
\label{gap_width_8}
\end{figure}

Fig. \ref{simulation_7} shows the trajectories of 15 ions from a 1 cm centered beam projected into the
x-y plane and x-z plane for the magnet whose field is
illustrated in Fig. \ref{Field_Map}. Notice that one of the ions fell out of the stable orbit. The orbits of the split sector design are much more sensitive to variations in the ions initial conditions as well as their effective magnetic rigidities, the profile of which is determined by the details of the charge exchange collisions. In a simulation with 1000 ions, this effect contributed to about 2$\%$ of losses. The rest of the losses were due to angular straggling or losses in the degrader. Losses in the degrader, however, were significantly reduced as compared to the previously investigated magnet design. Fig. \ref{gap_width_8}  shows the percentage of cooled ions as a function of the magnet gap
width for three different configurations; One with a 1 cm beam on
center, one with a 1 cm beam-spot offset by 2 cm, and another with
a 1 cm by 10 cm beam-spot offset by 5.5 cm along the
z-axis. Unlike previous simulations where analytical expressions
for the magnetic field were used, different pole gaps result in
different field profiles due to the lack of rotational symmetry
about the z-axis. It was therefore necessary to run a separate
simulation for each pole gap in order to take into account the
varying field profiles. As a result, Fig. \ref{gap_width_8} has a
limited number of data points compared to Figs. \ref{gap_width_5},
\ref{gap_width_6} $\&$ \ref{gap_width_7}.

\begin{figure}
\includegraphics*[scale = .45]{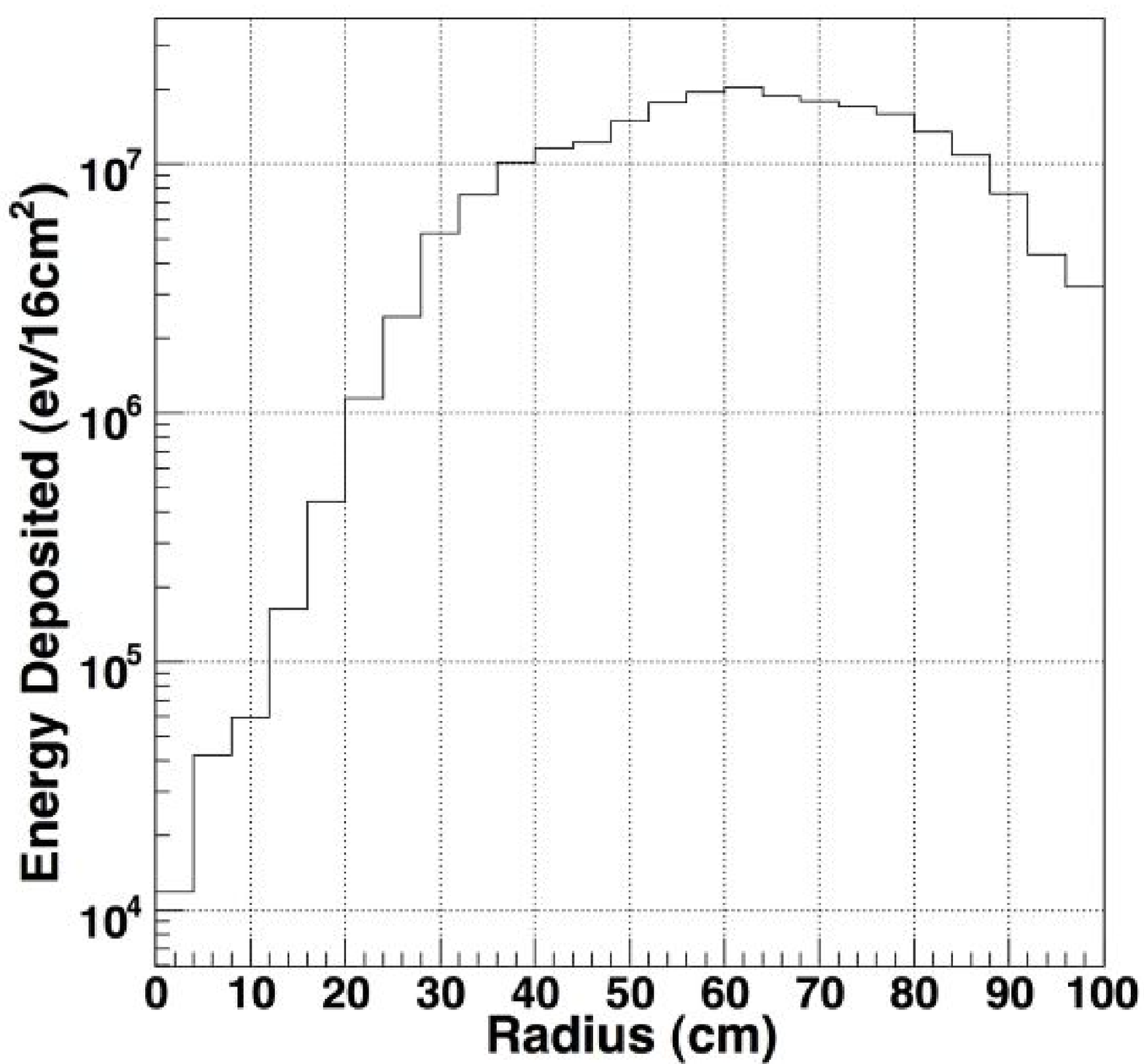}
\caption{Total energy deposition along radial direction for 50 bromine
ions of which only 28 were successfully cooled, plotted on a logarithmic scale in units of eV/(16 cm$^2$).}
\label{energy_dep_7}
\end{figure}

For centered beams the split sector design increases overall
efficiencies by reducing the loss of ions due to collisions with
the degrader, but the maximum achievable efficiency when the beams
are offset is about 85$\%$ for a 60 cm pole gap, roughly the same as the previously
studied magnets. The apparatus does, however, demonstrate
more effective focusing, which is evident by the gains in efficiency of
more than 35$\%$ for smaller pole gaps. The main drawback to this
method is the large extraction region and the increase in energy
deposited within this region. In this design ions move in elliptical orbits whose centers tend to drift away from the center of the magnet. In some cases the ions form highly eccentric orbits causing them to come very near the center of the magnet and depositing large amounts of energy within the extraction region during half of their rotation. While losing energy the orbits shrink about their center and in extreme cases some ions fail to make it back around the center of the magnet causing them to fall out of the system as seen in Fig. \ref{simulation_7}. Fig. \ref{energy_dep_7} shows
the energy deposited from a run with a 5 mm half width centered
beam using the magnet whose field is illustrated in Fig.
\ref{Field_Map}. In this run 28 out of 50 ions were successfully
cooled. The maximum density of deposited energy per cooled ion is found to be about 45 keV/cm$^2$, a  nearly 3 fold reduction compared to the simulation illustrated in Fig. \ref{energy_dep_3}. For the 30 cm gap used in this simulation, the resulting beam intensity limit is reduced by a factor of 2 to 3 from what was predicted in \cite{Bollen}. This, however, does not take into account that nearly half of the energy is deposited within the extraction region, where 7.5 GeV was deposited. This
corresponds to $6.5 \cdot 10^{6}$ ionized helium atoms per
cooled bromine ion. There is no doubt that this will place severe limits on extraction efficiencies and times as well as allowed beam intensities.

\section{Conclusion}

Extensive simulations of the cyclotron stopper have
been performed including physics processes that were either omitted or incorrectly treated in Ref.\cite{Bollen}. While it is difficult to place uncertainties on these calculations due to the small amount of experimental data available for the scattering of heavy ions in light gases, the results of these simulations agree well with the limited experimental data on gas filled spectrometers and charge-exchange cross-sections. These results are at variance with the simulations reported in Ref. \cite{Bollen}. We find the efficiency for the geometry proposed in that paper to be lowered by more than two orders of magnitude by the better treatment of the ion-gas interactions.

Increasing the pole gap of the magnet by a factor of about 15 allows one to recover a stopping efficiency of roughly 60$\%$. In addition, the factor of 15 increase in pole-gap required to handle the angular straggling allows the use of an offset degrader, which provides a further increase in the maximum achievable stopping efficiency approaching 90$\%$. Precise experimental determination of the radial straggling at low energies would be usefull to further refine these simulations. The newly proposed geometry results in a similar extraction time and a similar order of magnitude maximum intensity limit for the operation of this device. The final extraction efficiency, however, will likely be lower.

This paper, to maintain a direct comparison to Ref. \cite{Bollen}, only quotes stopping efficiencies. Not mentioned in both is the loss of ions due to neutralization in the stopping process. For stopping in He, several experiments suggest a maximum achievable neutralization survival probability \cite{Dendooven, Savard} during the process of slowing down in the range of 30-50$\%$ depending on the ion being cooled. Achieving efficiencies in this range is not a trivial task as it requires sub-ppb impurity levels within the stopping volume. Such efficiencies have only been accomplished by cryogenically freezing out any possible contaminants that may neutralize the ions (temperatures below 100 K were necessary to fully minimize the effects of neutralization in He) \cite{Dendooven} or through the use of ultra high purity helium in an ultra high vacuum system \cite{Savard}. In addition, there might also be losses due to inefficiencies in extraction that will arise at high intensity operation from the increased space-charge effects due to the large pole gap. For operation below the space charge limit, the overall efficiencies should be comparable to linear gas catchers, but with a much longer stopping range that is particularly useful for light ions. The main technical challenge with this approach will be the construction of such a large magnet geometry.

Finally, it has been found that the limited acceptance of the previously proposed apparatus
can be increased to cover a large fraction of the acceptance of modern
fragment separator with very little additional loss in efficiency by using an
offset degrader for this wider gap design. The space-charge handling capabilities for such a geometry are calculated to be about $10^{8}$ incoming ions per second, more than adequate for most applications.

\section{Acknowledgments}

This work was supported by the U.S. Department of Energy, Nuclear Physics Division, under contract No. DE-AC02-06CH11357.


\begin{thebibliography}{11}


\bibitem{ex1}
G. Savard, et al., Nucl. Inst. and Meth. B 204 (2003) 582.

\bibitem{ex2}
M. Wada et al., Nucl. Inst. and Meth. B 204 (2003) 570.

\bibitem{ex3}
G. Sikler, et al., Nucl. Inst. and Meth. B 204 (2003) 482.

\bibitem{ex4}
L. Weissman, et al., Nucl. Inst. and Meth. A 540 (2005) 245.

\bibitem{Limits1}
M. Huyse et al., Nucl. Inst. and Meth. B 187 (2002) 535.

\bibitem{Limits2}
A. Takamine et al., Rev. Sci. Inst. 76, 103503 (2005).

\bibitem{Katayama}
I. Katayama et al., Hyperfine Interactions 115 (1998) 165.

\bibitem{Bollen}
G. Bollen et al., Nucl. Inst. and Meth. A 550 (2005) 27.

\bibitem{SRIM}
J.F. Ziegler, SRIM-2003, http://www.SRIM.org/SRIM

\bibitem{Betz}
H.D. Betz, Rev. Mod. Phys. 44 (1972) 465.

\bibitem{SIMION}
Scientific Instrument Services, Inc., SIMION 7, http://www.SIMION.com

\bibitem{data1}
J. Alonso et al., Transactions on Nucl. Sci. Vol. NS-26, No. 3 (1979) 3686.

\bibitem{data2}
W.G. Graham et al., J. Phys. B: At. Mol. Phys. 18 (1985) 2503.

\bibitem{Paul}
M. Paul et al., Nucl. Inst. and Meth. A 277 (1989) 418.

\bibitem{Rehm}
K.E. Rehm et al., Nucl. Inst. and Meth. A 344 (1994) 614.

\bibitem{sternberginprep}
M. Sternberg et al., to be published.

\bibitem{Geissel}
H. Geissel et al., Nucl. Inst. and Meth. A 282 (1989) 247.

\bibitem{Weick}
H. Weick et al., Nucl. Inst. and Meth. B 164 (2000) 168.

\bibitem{Scheidenberger}
C. Scheidenberger, et al., Nucl. Inst. and Meth. B 204 (2003) 119.

\bibitem{fragRIA}
B.M. Sherrill, Nucl. Inst. and Meth. B 204 (2003) 765.

\bibitem{Dufour}
J.P. Dufour et al., Nucl. Inst. and Meth. A 248 (1986) 267.

\bibitem{Savard2}
G. Savard et al., Nucl. Inst. and Meth. B (2008), doi:10.1016/j.nimb.2008.05.091, in press.

\bibitem{EMIS07}
G. Savard et al., to be published.

\bibitem{Dendooven}
P. Dendooven et al., Nucl. Inst. and Meth. A 558 (2006) 580.

\bibitem{Savard}
G. Savard et al., Nucl. Inst. and Meth. B 204 (2003) 582.



\end{thebibliography}
\end{document}